\documentclass{article}

\usepackage{arxiv}

\usepackage[utf8]{inputenc} % allow utf-8 input
\usepackage[T1]{fontenc}    % use 8-bit T1 fonts
\usepackage{hyperref}       % hyperlinks
\usepackage{url}            % simple URL typesetting
\usepackage{booktabs}       % professional-quality tables
\usepackage{amsfonts}       % blackboard math symbols
\usepackage{nicefrac}       % compact symbols for 1/2, etc.
\usepackage{microtype}      % microtypography
\usepackage{lipsum}		% Can be removed after putting your text content
\usepackage{graphicx}
\usepackage[english]{babel}
\usepackage[square,numbers]{natbib}
\usepackage{doi}

\title{A Review on Searchable Encryption Functionality and the Evaluation of Homomorphic Encryption}

\author{ \href{https://orcid.org/0000-0000-0000-0000}{\includegraphics[scale=0.06]{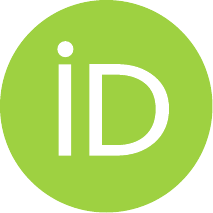}\hspace{1mm}Brian Kishiyama} \\
	Department of computing, engineering and mathematical sciences\\
	Texas A\&M University - San Antonio\\
	San Antonio, TX 78224 \\
	\texttt{bkish01@jaguar.tamu.edu} \\
	\And
	\href{https://orcid.org/0000-0000-0000-0000}{\includegraphics[scale=0.06]{orcid.pdf}\hspace{1mm}Izzat Alsmadi} \\
	Department of computing, engineering and mathematical sciences\\
	Texas A\&M University - San Antonio\\
	San Antonio, TX 78224 \\
	\texttt{ialsmadi@tamusa.edu} \\
}

\renewcommand{\headeright}{Technical Report}
\renewcommand{\undertitle}{Technical Report}
\renewcommand{\shorttitle}{Information Security}

\hypersetup{
pdftitle={A Review on Searchable Encryption Functionality and the Evaluation of Homomorphic Encryption},
pdfsubject={q-bio.NC, q-bio.QM},
pdfauthor={Brian~Kishiyama, Izzat~Almadi},
pdfkeywords={Security Information, Privacy, Searchable Encryption, Homomorphic Encryption, Cloud Computing},
}

\begin{document}
\maketitle

\begin{abstract}
	Cloud Service Providers, such as Google Cloud Platform, Microsoft Azure, or Amazon Web Services, offer continuously evolving cloud services. It is a growing industry. Businesses, such as Netflix and PayPal, rely on the Cloud for data storage, computing power, and other services. For businesses, the cloud reduces costs, provides flexibility, and allows for growth. However, there are security and privacy concerns regarding the Cloud. Because Cloud services are accessed through the internet, hackers and attackers could possibly access the servers from anywhere. To protect data in the Cloud, it should be encrypted before it is uploaded, it should be protected in storage and also in transit. On the other hand, data owners may need to access their encrypted data.  It may also need to be altered, updated, deleted, read, searched, or shared with others. If data is decrypted in the Cloud, sensitive data is exposed and could be exposed and misused. One solution is to leave the data in its encrypted form and use Searchable Encryption (SE) which operates on encrypted data. The functionality of SE has improved since its inception and research continues to explore ways to improve SE. This paper reviews the functionality of Searchable Encryption, mostly related to Cloud services, in the years 2019 to 2023, and evaluates one of its scheme, Fully Homomorphic Encryption. Overall, it seems that research is at the point where SE efficiency is increased as multiple functionalities are aggregated and tested. 
\end{abstract}

% keywords can be removed

\section{Introduction}
Examples of some of the resources or services provided in cloud computing are dedicated virtual machines, core computing, security and identity, virtual networks, machine learning platforms, cloud storage, and streaming analytics \cite{cloud_compare_2023}. The cloud provides services to Location Based Services (LBS), Internet of Things (IoTs), electronic Health (eHealth) research and organizations, blockchains, smart devices, and other fields. Data sent to the cloud may contain sensitive information. This includes personal or customer information, user location and geographical habits, medical records, and trade secrets. Businesses, such as Netflix, Zoom, and PayPal, rely on the Cloud as part of their business structure. The Cloud is accessed by the Internet which makes it a target for hackers. It can also be a target from a business insider looking to make money off of a company's trade secret.  Data such as personal or financial data is a precious resource that needs protection.  One method of protecting data is to encrypt the data in storage and transit. At times, however, the encrypted data needs to be operated on. For example, the data would need to be decrypted in order to search for a file or to change a record. If this data is decrypted, plain-text is exposed to potential adversaries. So, the data should remain encrypted in the cloud while in transit to other locations, at rest in memory, or when used. 
 
A proposed solution is to keep data encrypted while in the cloud. When needed, and decryption needs to occur, transport the encrypted data out of the cloud and back to the data owner. The owner decrypts the data in his or her protected private server. Changes are made on plain-text data. After changes are made, the owner encrypts the data and then sends it back to the cloud. 

The problem with this solution involves the download of the entire database, decrypting the data, operating on the plain-text, re-encryption, and then sending it back to the cloud. This is not worthwhile. This would not only involve time but would require memory to hold the data, high computational costs, and heavy network usage. 

A recently proposed solution is to utilize Searchable Encryption (SE).  With this method, data in the cloud or other server is never decrypted.  It remains encrypted, even when operated on \cite{dawn_xiaoding_song_practical_2000}. SE allows searches, updates, or calculations on encrypted data such that its plaintext is never exposed.

There are different SE schemes that allow operations on encrypted data. Data is never decrypted. One type of scheme is Homomorphic Encryption (HE). Its definition meets the definition of SE such that it operates on encrypted data while stored in an untrusted server. Formally, "homomorphic encryption systems allow data to be analyzed and processed on a ciphertext rather than the underlying data itself" \cite{ieee_what_2021}. HE distinguishes itself from SE such that HE can add or multiply numbers while in encrypted form.  An issue with HE is that it involves heavy memory and computation costs.  Some research indicates it may not always be practical. 
 
In fact, there are concerns from online articles. For instance, one article claims that the HE process, when compared to data that is not encrypted, is one million times slower when conducting operations on data \cite{cutress_intel_2021}. This article, \emph{Intel to Build Silicon for Homomorphic Encryption}, talks about a complex dataset such as a Magnetic Resolution Image (MRI) that undergoes screening and processing. Dr Cutress claims that homomorphic encryption is not a viable option with MRIs. He claims that one second of plaintext processing is equivalent to a week of HE processing. So, it appears that HE is limited in its use when dealing with complex datasets. In terms of research contribution, however, there are hundreds of research projects on the subject. Google Scholar produced over 51,000 articles related to homomorphic encryption. 

Our paper reviews the functionality of Searchable Encryption, which includes Homomorphic Encryption, from 2019 to 2023. 

\subsection{Cloud and Security}

The Cloud is widespread and a massive infrastructure. It stores and retrieves data. Privacy of sensitive data is an issue \cite{amorim_leveraging_2023}.  Encryption schemes have been developed to ensure privacy. Organizations as well as individuals use the cloud to store data.

An example of Cloud usage is when airports use Cloud Services as part of their infrastructure. This technology increases profits. Manual labor is shifted to automated operations in the Cloud. In addition, airport networks integrate with IoT, sensors, GPS, surveillance cameras, facial recognition, etc. Data stored in the Cloud contain trade secrets, as well as customer information, that the airlines rely on. The airline industry, with its data and operations in the Cloud, is vulnerable to cyber attacks \cite{malik_homomorphic_2023}. 

Netwrix claim that 80\% of cloud users store sensitive data \cite{netwrix_company_about_2023}. Sensitive data includes employee and customer data, financial records, and trade secrets. Attacks on sensitive data include phishing, ransomware, data leakage, and data loss or theft. Businesses surveyed report that data breaches can be costly and leads to unplanned expenses. 

In regards to healthcare, according to Netwrix, 55\% of healthcare organizations have experienced a data breach. If a data breach occurs, the healthcare company will need to protect the consumer. One method used is to offer the consumer a credit monitoring service, such as IDX. 

IDX is a data breach response company.  In their research, they found that data breaches include "stolen or compromised credentials (19\%), phishing (16\%), cloud mis-configuration (15\%), vulnerabilities in third-party software (13\%), and malicious insiders (11\%)" \cite{idx_cost_2022}.  They further claim, via an IBM survey, that 83\% of surveyed businesses reported at least one data breach. Companies identified as "organizations in critical infrastructures", e.g., banks and technology companies, have an average expense of \$4.82 million. This cost can go up or down depending on the security defenses used \cite{idx_data_2023}.    

Overall, there are advantages of using the cloud. They include data availability, reduced business costs, and unlimited storage space for data. However, they are not immune from unauthorized access, data breaches or insider threats \cite{amorim_leveraging_2023}.    

\subsection{Security Information}
Encryption is needed to protect sensitive data. There are many encryption schemes such as secure search, private information retrieval (PIR), and searchable encryption (SE) \cite{amorim_leveraging_2023}.  When protecting data, there are three different areas. The first is data "in transit," such as data traveling across the Internet. Second, there is data "at rest," or simply data in storage or sitting in memory. Finally, there is data "in use." This is when data is created, read, updated, deleted, searched for, added, or any other operation. Data needs protection in all of these three states.

\section{Searchable Encryption}

Searchable encryption has many uses as it allows operations on encrypted data. For instance, a search operation is conducted on encrypted data without the decryption to plaintext. Because the data remains encrypted, it can be used to secure data in the Cloud while at rest, in transit, and in use. 

Searchable encryption appeared in 2000. It performs search operations while maintaining privacy and confidentiality. Song et al. proposed that untrusted servers should only have encrypted data, i.e., data should not be decrypted while in the server. Although some functionality is lost, data remains encrypted even when searching for words \cite{dawn_xiaoding_song_practical_2000}. This makes sense since storing data on servers, such as email, in its encrypted form makes data unreadable. In turn, it reduces security and privacy risks. If an attacker gained access to the server, he or she would only see ciphertext. Moreover, the server remains unaware of what is stored, even when conducting searches. In SE schemes, the server is never given plaintext; the server is not given keys to decrypt the data - for any operation. To ensure that a server never sees plaintext data, the data owner encrypts the data before uploading to the server.   

Users have different options to protect their data. They may use several keys at different locations to encrypt documents. Stronger schemes include the changing of encryption keys, re-encrypting all documents, and re-ordering ciphertext \cite{dawn_xiaoding_song_practical_2000}. These methods are used to ensure an outsourced or untrusted server cannot deduce the meaning of words. 

Normally, searchable encryption involves five steps \cite{Dai_X_2019}:
\begin{enumerate}
    \item Extract the dataset features such as keywords from the documents. Most schemes use TF-IDF to extract the features, as explained below.
   
    \item Build a secure index with keywords.
    \item Encrypt the dataset.
    \item Generate a search trapdoor, i.e., an encrypted query.
    \item Search the index and return the results.
\end{enumerate}

As explained later, there are various methods to conduct searches such as using single or multiple keywords. Current research involves multi-keyword searches such as the Multi-Keyword Rank Searchable Encryption (MRSE) scheme based on Homomorphic Encryption. This uses the Paillier cryptosystem with Threshold Decryption (PCTD) algorithm \cite{yang_multi-user_2020}, an indexed based search scheme.

\subsection{Searchable Symmetric Encryption}
Searchable Symmetric Encryption (SSE) is an example of a SE scheme that is a symmetrical type.  It uses a symmetric key to encrypt data and allows other authorized users to search the encrypted data using certain keywords or search terms.  After the search is conducted, results are returned in encrypted form \cite{Sucharitha_2023}. The authorized user would use his or her key to decrypt the message.

\subsection{Public Key Searchable Encryption}
Public Key Searchable Encryption (PKSE) is another popular type of searchable encryption scheme. This is an asymmetric type of encryption scheme. The public key is used to encrypt both the data and keywords. The encrypted keywords are sent to the server. The server uses its private key to search the encrypted data and return the results. The results are returned back to the user in encrypted form \cite{Sucharitha_2023}.    

\subsection{Attribute-Based Encryption}
Another scheme is the Attribute Based Encryption (ABE) scheme that uses fuzzy identification to implement fine-grained data access control. This is needed since data owners cannot invoke access controls on outsourced data . ABE can be used where cloud servers perform complex computing tasks \cite{Sucharitha_2023}. Additionally, it allows encrypted data to be searched with searchable encryption algorithms, allows file updating, and attribute revocations \cite{Lin_2023}. 

Simply, ABE makes fine-grained access control manageable by allowing specified persons to access certain encrypted files.

\subsection{Order-Preserving Encryption}
Order-Preserving Encryption (OPE) is an SE scheme that allows comparisons to occur over encrypted data. OPE allows the ciphertexts to maintain the same order as its plaintext version. So, if plaintext 1 is greater than plaintext 2, then Encrypted(plaintext 1) is greater than Encrypted(plaintext 2).  With this scheme, servers can perform comparisons, ordering, ranking, and range queries over encrypted data \cite{Shen_A_Prac_2021}. 

\subsection{Homomorphic Encryption}

An encryption scheme is Homomorphic Encryption (HE). It allows operations on the encrypted data without the need to decrypt them \cite{Pisa_Global_2012}. Homomorphic Encryption is a type of searchable encryption. Of the SE schemes, Fully HE is a relatively new scheme when compared to the others. HE is the encryption of data and then analyzing and processing the ciphertext while never exposing it in decrypted form \cite{ieee_what_2021}. Fully HE allows unlimited mathematical operations, primarily addition and multiplication, on the encrypted data. Because the data remains private and plaintext is not shown, outside parties are unable to understand the data \cite{gillis_what_2022}.  

\subsection{Types of Homomorphic Encryption}
Research has expanded Homomorphic Encryption, and developed variation. However, there are three common types. This includes Partial HE, Somewhat HE, and Fully HE.

\subsubsection{Partially Homomorphic Encryption (PHE)}
Of the three common HE schemes, PHE is the most widely used \cite{amorim_leveraging_2023}. It is also the least secure. This has been around since 1978 \cite{ieee_what_2021}. Rivest, Adelman, and Dertouzos proposed it as "privacy homomorphisms." They proposed that data remains encrypted and is never decrypted. They found that some functions exist to allow operations to be performed while the data is encrypted. The only time that it is decrypted is when it is downloaded to the user's computer \cite{rivest_data_1978}.

PHE is efficient over both SHE and FHE, but PHE is limited to one type of math operation, either addition or multiplication. 

It is still in use and researched. For example, a recent study proposed that PHE be used at airports to secure data, or keep it private, while in the Cloud \cite{malik_homomorphic_2023}. The scheme uses a Single Keyword Searchable Encryption (SKSE) scheme based on the Paillier Cryptosystem.

\subsubsection{Somewhat Homomorphic Encryption (SHE)}
This also has been around since 1978 \cite{ieee_what_2021}. SHE can perform both addition and multiplication. With SHE, the number of homomorphic operations are limited \cite{Pisa_Global_2012}. As more operations are performed, more noise or error, is introduced in the calculations. At some point, too much error causes loss in accuracy and decryption is not possible. SHE is not as efficient as PHE, but it is more efficient than Fully Homomorphic Encryption (FHE). Because SHE is faster and more compact than FHE, it is useful to more applications \cite{yasuda_new_2015}. 

\subsubsection{Fully Homomorphic Encryption (FHE)}
Fully Homomorphic Encryption (FHE) is the third common type of HE schemes. FHE is relatively new in comparison to other techniques. Regarding FHE, a popular paper is Dr. Craig Gentry's FHE scheme, proposed in 2009. His scheme leads to addition and multiplication on encrypted ciphertext an unlimited number of times \cite{Gentry_2009}. Dr. Gentry's scheme basically controls the size of math operations on ciphertext so they do not grow uncontrollably large. Although, research has noted the high computational costs and large ciphertexts involved in FHE.  

In Dr. Gentry's scheme, bootstrapping is utilized, i.e., he uses a decryption algorithm on a produced ciphertext, along with the private key, after each operation. This reduces the "noise" and allows infinite homomorphic operations \cite{Pisa_Global_2012}. The decrypted ciphertext then gets re-encrypted. Without the reduction, the noise grows and accumulates after each operation. If it becomes too large, the ciphertext cannot be decrypted.  

Noise is a "bounded element of randomness in the encryption scheme added to the message so that it is easy to remove and unveil the original message for trusted parties and conceal the message from adversaries" \cite{hovd_handling_2017}.   

Nowadays, FHE is combined or modified and used in encryption schemes. One example is where Rajan et al. uses dynamic multi-keyword search that uses a modified fully homomorphic encryption scheme along with Prim's algorithm to increase efficiency \cite{rajan_dynamic_2019}.

\subsection{FHE Open Source Libraries}
There are many Fully Homomorphic Schemes. According to Wikipedia, there are 12 open source libraries for FHE \cite{wikipedia_homomorphic_2023}:
\begin{itemize}
    \item HElib. This is developed by IBM. It includes the BGV and CKKS schemes
    \item Microsoft SEAL: Its developer is Microsoft and includes the BGV, CKKS, and BFV schemes
    \item OpenFHE: These has several developers that include, Intel, MIT, Samsung Institute of Technology, UC San Diego, and others. This is the successor to Palisade and includes all schemes, i.e., BGV, CKKS, BFV, FHEW, CKKS Bootstrapping, and TFHE.  
    \item PALISADE: The developers include New Jersey Institute of Technology, MIT, UC San Diege, and others. Like OpenFHE, it includes all schemes except CKKS Bootstraping.
    \item HEAAN: Developed out of Seoul National University. Its schemes include CKKS and CKKS Bootstrapping.
    \item FHEW: It has two developers, Leo Ducas and Daniele Micciancio and focus is on the FHEW scheme.
    \item TFHE: It has a team of developers, Chilotti et al. with a focus on the TFHE scheme.
    \item FV-NFLlib. Developed by CryptoExperts with a focus on BFV.
    \item NuFHE: The developer is NuCypher with a focus on FHE.
    \item REDcuFHE: The developer is TwC Group with a focus on FHE.
    \item Lattigo: The developers are EPFL-LDS and Tune Insight. Both EPFL-LDS and Tune Insight conduct research on data security.   
    \item TFHE-rs: The developer is a company called Zama. They built applications using HE, with a focus on FHE.
\end{itemize}

Examples of other platforms that exist are cuHe, Concrete, and EVA,  \cite{he_standarization_introduction_2018}.

Above all, there are three "effective and peer-reviewed FHE schemes" which are Brakerski-Gentry-Vaikuntanatha (BGV), Brakerski, Fan-Vercauteren (BFV), and Cheon-Kim-Kim-Song (CKKS). With the various schemes and developments, they should not be implemented unless the user understand them. Some HE platforms are emerging and not fully developed \cite{dilmegani_what_2023}. Some schemes have emerged but not yet been fully tested, and some schemes are effective but are proprietary \cite{ieee_what_2021}.  

\subsection{Advantages of Homomorphic Encryption}
Cloud services can benefit from HE as it preserves privacy and companies can share data without compromising sensitive information. At this time, it is impenetrable by quantum computers \cite{ieee_what_2021}.   

Businesses use cloud services and storage services and may have security concerns. HE is a promising solution. The cloud when used with HE allows businesses to share sensitive information with trusted parties. HE also helps with regulations where privacy, such as in areas of finance and medical records, is required by law. 

\subsection{Disadvantages of Homomorphic Encryption}
The database is an asset for most organizations and needs to be protected. Protection includes multiple layers of security. This includes firewalls, authentication, access control, and database encryption. Database Encryption has two disadvantages  The first is \textit{Key management}. Authorized users must have access to the decryption key if they need to access the encrypted data.  Because a database is typically accessible to a wide range of users and a number of applications, providing secure keys to selected parts of the database to authorized users and applications is a complex task \cite{stallings_computer_2018}.

The second disadvantage of an encrypted database is \textit{inflexibility}. This refers to parts or all of an encrypted database. Encryption can be applied to the entire database, at the record level (encrypt selected records), at the attribute level (encrypt selected columns), or at the level of the individual field. Performing record searches on encrypted data is a highly complex task where not much flexibility is given.

Another disadvantage, is that encryptions such as FHE is too slow to be useful. FHE is claimed to be too slow to have practical applications.  It is about 1 million times slower than if operations were in plaintext \cite{cutress_intel_2021}. This, however, appears to focus on complex datasets.

\section{Schemes and Tools used to Improve Searching Functionality}
When Song et al. proposed their paper in 2000, they used a sequential scan search.  Each document contains a set of n-bit keywords. Each keyword gets encrypted. This provides a document full of encrypted keywords. When a search is conducted, it looks for the encrypted keyword. So, a search is done on all documents. Documents that contain the encrypted keyword are returned. The keywords are never decrypted to plaintext unless downloaded to the data owner with the decryption key. Thus, the server does not know what is being searched for. There are limitations with the sequential scan \cite{sharma_searchable_2023}:

\begin{enumerate}
\item Searching is limited to a single keyword search.
\item The search complexity is linear which depends on the number of documents and the size of the documents.
\item The encryption of the general files cannot be applied like the encryption of the documents.
\item The files cannot be compressed to save space. If compressed, the documents cannot be searched.
\item The files are vulnerable to statistical attacks. 
\end{enumerate}

Despite the initial limitations, research has continued to improve efficiency or the functionality of SE. This includes single keyword searches, multi-keyword searches, range searches, fuzzy keyword searches, verifiable searches, image searches, multiple users searches, bloom filter searches, etc.

\subsection{Single Keyword}
Symmetrical Searchable Encryption was proposed by Song et al \cite{dawn_xiaoding_song_practical_2000}. In this scheme, a document in a cloud server is searched by progressing through the whole document. A single, specific word is used to determine if it is in the document. Lately, single keyword is not researched unless search efficiency is demonstrated. Using single keywords has limited functionality. It may be researched with a particular scheme such as Partial Homomoprhic Encryption that uses Paillier Cryptosystem \cite{malik_homomorphic_2023}. Overall, many schemes may use single keywords for searching; after all, it is more efficient when compared to multi-keyword searches. 

\subsection{Multiple Keywords}
Using multiple keywords to search an encrypted database is needed since it provides better functionality with today's applications. When moving from a one keyword search to a multi-keyword search, there is a noticeable drop in efficiency \cite{tahir_parallelized_2020}. So, attempts are made to reduce the computation overhead and increase efficiency \cite{Cui_Y_2020}, or make searching faster \cite{Liu_G_2022_FASE}. In addition, research increases functionality by using multi-keyword searches alongside other functionalities. For example, it can be combined with fuzzy and semantic searches \cite{Dai_X_2019}. Multi-keyword searches can use ranked, searchable ABE schemes (FEMRSABE) \cite{Lin_2023}. Moreover, multi-keyword searches increases its functionality by including research to improve its security. For instance, the Multi-Keyword Ranked Search Scheme with Fine-grained access control (MRSF) provides multiple keyword searches along with access control in a cloud environment that maintains encrypted data \cite{Li_J_Prac_2022}. 

Research is also extended to "ranking" as in the multi-keyword ranked searchable encryption (MRSE).  In this scheme, multi keywords are used in a query and the top search results are returned. This is a nice feature as we only want the most relevant documents returned to us \cite{liu_multi-keyword_2021} \cite{Shen_2021}. There is another multi-functionality that this scheme provides. If a user does not use the correct keyword for searching, the server may infer another keyword \cite{liu_multi-keyword_2021}. Other combined functionalities include the searching of multi-keywords and by allowing files to be updated \cite{hussain_blockchain-based_2022}, \cite{ashwini_productive_2022}.  Finally, it can be combined and improved with index-based platforms \cite{srivani_multi-key_2022}, \cite{ashwini_productive_2022}.  

\subsection{Ranges and Spatial Queries}
Ranges are different from using multiple keywords. Ranges could involve a specifying a keyword and a number. A user specifies the number of times a keyword appears in a document in order for it to be returned \cite{amorim_leveraging_2023}.  Ranges could also include spatial attributes or geographical keyword range queries over encrypted data \cite{Gong_Effi_2023}. Geometric range queries are useful in research as it can find data points within an area such as a rectangle or circle \cite{li_efficient_2019}. This is useful, as related to location based services, to not to reveal a person's location when looking for a service or Point of Interest. It also extends to Internet of Things (IoT), vehicle networks, valet parking, and other services \cite{Song_Priv_2022}.

\subsection{Fuzzy Keyword Searching}
Most schemes only support exact keyword searches. If a word is misspelled, the user will not be able to retrieve any records or files from the server \cite{Li_2023_VRFMS}. Fuzzy word resolves this issue but the searches increase computational costs. Research has improved to allow more misspellings and to increase the search efficiency \cite{li_multi-keyword_2021}. Like the other functionalities, fuzzy keyword searches extends, or combines, with other functionalities such as multi-keywords, phrases, ranking, blockchains, semantics, etc. 

\subsection{Ranked Searches and TF-IDF}
\begin{quote}
    "TF-IDF is a statistical method used to evaluate the importance of a word in a file collection...There is a hypothesis that the most meaningful words for distinguishing documents should be those that appear frequently in a single document and less frequently in other documents in the entire file collection" \cite{Li_2023_VRFMS}. 
\end{quote}

Ranked searches allow the most relevant documents to be returned based on the keywords used \cite{Shen_2021}.  Although I did not specifically track "TF-IDF," it appears often in the articles that I searched. It is worthwhile to understand and explain as it relates to ranked searches. It is a common method used in finding the most relevant keywords \cite{ali_efficient_2023} \cite{Lin_2023}, and widely used and integrated in index development and query generation \cite{ashwini_productive_2022}.  

TF stands for Term Frequency and IDF stands for Inverse Document Frequency.  In mathematical terms \cite{geeks_understanding_2021}:

\begin{itemize}
    \item tf-idf(t, d) = tf(t, d) * idf(t)
    \begin{itemize}
        \item tf(t, d) = t / d
        \begin{itemize}
            \item the number of times that the word 't' occurs in a document.
            \item 'd' is the total number of words in the document.
        \end{itemize}
    \end{itemize}
    \begin{itemize}
        \item idf(t) = ${log}_{2}$(N/df(t)) 
        \begin{itemize}
            \item 'N' is the total number of documents in a collection.
            \item df(t) is the number of documents that contain the word 't'.
        \end{itemize}
    \end{itemize}
\end{itemize}

Research continues with ranking and retrieval by itself as the efficiency may be improved \cite{ahmad_coeus_2021}. Ranked searches are used in combination of other functionalities such as multi-keyword searching. It is not uncommon to see "multi-keyword" and "ranked" to be paired. "Multi-keyword ranked search" is used quite often since it quickly returns the most relevant documents to the requester \cite{Shen_2021}. Multi-keyword ranked searches can be paired with security for access control and analysis \cite{Li_J_Prac_2022}. 

\subsection{Indexes}
Indexes are search structures that are researched to improve the search performance. A problem can arise, especially in big datasets, where the index becomes the "processing bottleneck" \cite{Zobaed_ClustCrypt_2019}.

Index searches are in a different category from the sequential scan searches. Some index structures conduct sequential searches with different types. According to Amorin et al., index structures include simple indices, inverted indices, and tree indices. Index searches reduce the number of comparisons made and are efficient, but queries are limited to keywords contained in the index \cite{amorim_leveraging_2023}.  

A simple index structure involves two  actors, the data owner and the cloud service provider. The data owner sends his encrypted data to the cloud service provider, along with an index of keywords. When the owner wants a document, he sends a search token - with a keyword - to the cloud service provider. The cloud service provider uses the search token to locate the keyword and retrieve the document \cite{yan_secure_2023}. 

Normally, an index is when a data owner extracts keywords from plaintext files. The extracted keywords are used to build a "Secure Index." The plaintext is then encrypted with symmetric encryption. The keywords are then transformed into trapdoors. The cloud servers that store the encrypted data match an authorized user's trapdoor to to the Secure Index.  This produces search results of the user's target keywords, which are returned to the user. This method is used to improve efficiency and security \cite{Lin_2023}.    

Indexes can be expressed as search trees.

\subsection{Search Trees}
\begin{quote}
    "Several recent schemes employ a server-side encrypted index in the form of a search tree where each node stores a bit vector denoting for each keyword whether any file in its sub-tree contains that keyword...the way data is distributed in such a search tree has a big impact on the cost of searches" \cite{etemad_optimizing_2018}.
\end{quote}

\textit{Tree-based searching} methods provide range searches. These are good for numerical datasets, and may be used in other search schemes such as FHE schemes \cite{Lin_2023}.

There are different types of trees that range from simple to complex and can be used in search schemes. Trees split data or data structures and provide choices or decisions to access data. For instance, the binary decision (BD) tree has a dataset in the parent node and splits its attributes into the child nodes. The splitting continues until the leaves are reached. The leaf nodes are subsets of the attribute set \cite{Zhang_KMSQ_2023}. 

For complexity, trees can be reshaped and used in different ways. For example, a segment tree is a "binary tree-based data structure for conducting efficient range aggregation queries on a sequential data set" \cite{Guan_Toward_2021}. In this tree structure, the inner nodes stores the results of range aggregation function and the leaf nodes store a data record from the data set.  

Search trees provide only part of the functionality of an encryption schemes and are combined with other functionalities. For instance, by using a search tree data structure, Lin et al. created two different flexible and efficient multi-keyword, ranked searchable attribute-based schemes (FEMRSABE) \cite{Lin_2023}. These tree schemes use fuzzy keyword searching as well as semantic searches. 

Efficiency may be improved with certain methods of how trees are searched. There are different methods such as Depth First Search, Breadth First Search, Boundary Traversal, Post Order Traversal etc. For example, Bhavya et al. used the Greedy Depth-first Search in their proposed Efficient File Upload and Multi-Keyword Search over Encrypted Cloud Data (EFUMS) scheme \cite{Bhavya_Conf_2020}.  In Ashwini's research, he also proposed the Greedy Depth-first Search in his tree for his multi-keyword, ranked scheme to improve efficiency \cite{ashwini_productive_2022}. 

Efficiency is also increased depending on the type of query. Li et al. proposed a R-Tree index for their geometric range query scheme on encrypted data \cite{li_efficient_2019}. R-Trees are efficient for indexing spatial data although B+ trees are the most common structures in databases. B+ trees can index data in single, sequential order; R trees store multi dimensional data which makes them better for range scans \cite{sypykowski_r-tree_2022}. 

When searching for keywords, indexed in a tree, the tree can be an elaborate structure such as the Four Branch Tree \cite{Wu_Four_2019}. The structure uses a keyword binary trees. Each node in the tree contains a set of keywords. Although, the nodes are allowed to be empty. The structure also implements a w-tree, which is a sub-type of a keyword tree. Each node, at most, contains "w" elements inside. Meaning, if a keyword contains \{a,b\}, \{a\}, or \{b\} in its parent node, then a w-tree can be split into two trees such that one tree contains all a's in its nodes and the other split tree contains all b's in its nodes.   

\subsubsection{Searching Graphs}
Graphs have vertices and the vertices are connected by edges. Graphs, between two connected vertices, can flow in one direction, i.e., directed graphs, or flow in both directions, i.e. undirected graphs. 

There are similarities and differences between graphs and trees. Graphs may or may not have leaf nodes. Trees, in comparison to graphs, have leaf nodes, where every leaf can be access from a root node.  Although, some graphs can have a root node.  A tree is considered to be a special type of graph, but acyclic and always directed. Meaning, graphs can loop but trees never do. Trees flow in one direction while graphs can flow in one or both directions. 

Researchers are not bound when formulating graphs or trees when improving the functionality. Improvements can be made in a "rooted hierarchical graph structure" \cite{bingu_security_2022}. Searching graphs while the structure is encrypted is resource intensive. Nevertheless, research continues to improve its efficiency. For instance, Ge et al. proposes a matching query scheme that supports subgraph extraction to improve efficiency\cite{Ge_Privacy_2023}. Their scheme uses two cloud servers for the matching operation yet ensures privacy. In additional to efficiency, graph searching can be combined with other functionalities, like the other functionalities previously addressed. For example, while using road networks, where graphs are often used, misspelling can be rectified \cite{sun_constrained_2021}. Encrypted graph searching is not limited to road networks and seen in different schemes. For instance, a graph is applied in Industrial Internet of Things (IIoT).  IIoT refers to end devices that uses sensors, networking, information gathering and processing, security, and efficiency in an business or industrial environment \cite{Ge_Transactions_2022}.

\subsection{Verifiable Searching}
\begin{quote}
    "Untrusted cloud servers may return incorrect or incomplete results...Since cloud servers are mostly untrusted, verifiable search has become a hot topic among scholars" \cite{tang_vr-peks_2023}.
\end{quote}

Data in the cloud is susceptible to attacks. Encryption ensures confidentiality but what if an attacker alters our encrypted data. It is important to ensure our data is trustworthy by checking its integrity. Verifiable SE schemes can verify if the returned results have be modified or deleted.  This uses high overhead \cite{Li_2023_VRFMS}. Verifiable schemes, verifies the results, but also allows all involved parties to monitor the legitimacy of others \cite{yan_secure_2023}. Depending on the scheme, verification could be improved. Some schemes may not be as effective since some verifiable results are not "comprehensive"; they need to demonstrate correctness and integrity \cite{duan_verifiable_2023}.

Like the other functionalities, research has combined verification with other searchable functions. 

\subsection{Image Searching}
\begin{quote}
    "For security and privacy concerns, images (e.g., medical diagnosis, personal photos) should be encrypted before being outsourced. However, traditional encrypted image retrieval techniques still suffer from costly access control and low search accuracy" \cite{Li_Trace_2022}.
\end{quote}
Images include personal photographs and medical images that need protection. They also include high value images that need to be protected from copyright infringement. Some people make money from images and want to control their use \cite{widmer_what_2017}. Images can be encrypted and later searched for.  Retrieval of images of some encryption schemes have low accuracy return \cite{Li_DVREI_2022} \cite{Li_Trace_2022} \cite{Krishna_secure_2023}.  With image searching, efficiency or speed along with accuracy seems to be the focus of most of the research in this area.

\subsection{Multiple Users}
Not all platforms use single users. Some platforms will need multiple users to access the encrypted data. Especially in the medical field where collaboration, data sharing, and collective intelligence is needed for medical research \cite{ali_efficient_2023}. This is an important functionality since many servers are semi-trusted yet multiple users rely on cloud storage solutions to protect personal data \cite{Ye_AVeri_2019}. 

\subsection{Bloom Filters}
Bloom Filters provide an important tool in SE. There are different ways of searching data, such as linear searches or binary searches. In contrast, a Bloom Filter is a different type of search method. "A bloom filter is a space-efficient probabilistic data structure that is used to test whether an element is a member of a set" \cite{g2g_bloom_2017}. It uses hashing where hashing takes an input, and outputs a unique identifier of fixed length. This unique identifier identifies a member of a set.

According to Liang et al., SE schemes that use bloom filters for searching are called \textbf{Bloom Filter-based Searchable Encryption}. Their study further claims that these searches are highly efficient over other schemes without bloom filters \cite{Liang_PrivBloom_2023}. In their study, the Bloom-filter based SE scheme is improved upon.

\subsection{Other Functionalities}
There are several other functionalities that have been improved but not detailed and listed that involve encrypted data use. This includes semantics, words with similar meanings, \cite{Lin_2023}, multiple data owners with multiple users \cite{wang_secure_2019}, deduplication of data or the removal of duplicated data \cite{swathika_time-conserving_2023}, comparing multiple dimensional data between two different users without revealing the users' data \cite{Shen_A_Prac_2021}, having the ability to query encrypted data sets with heterogeneous data sources \cite{Feng_Trans_2022}, clustering encrypted unstructured big data \cite{Zobaed_ClustCrypt_2019}, etc. 

\subsection{Multiple Functionalities}
Searchable encryption is not limited to improving one type of function. They mostly include multiple functionalities. For instance the Verifiable Ranked Fuzzy Multi-keyword Search scheme (VRFMS) uses a hashing and bloom filter. As you can see, it utilizes fuzzy keyword searches that also employs multiple keywords, ranking, a bloom filter, and a verifiable search scheme \cite{Li_2023_VRFMS}. 

As mentioned in Search Trees, above, the FEMRSABE also include various functions that have a wide range of functionalities. 

Overall, this paper looks at studies within the last five years. It seems that the functionality of SE continues to improve in different areas. We also see the different searchable encryptions schemes. This is important to understand as research continues to expand functionality of schemes. 

\subsection{Understanding Homomorphic Encryption}
The various homomorphic schemes are complex and part of the Searchable Encryption family. Paillier Homomorphic Encryption is a partial homomorphic encryption (PHE) scheme while Pyfhel is a Fully Homomorphic Encryption (FHE). FHE schemes being the latest of the Searchable Encryption family. 

We look at Paillier's HE as demonstrated on Wikipedia.  It is public and accessible with ample resources explaining the process. We also look at open source Pyfhel since it is offered in Python and its GitHub files have been recently updated and maintained.

\subsubsection{Paillier's HE}
Partial Homomorphic Encryption (PHE) schemes are the most widely used.  In 1999, Pascal Paillier invented the Paillier Cryptosystem \cite{wikipedia_paillier_2023}. It is an asymmetric homomorphic encryption system that can add two ciphertexts or allows scalar multiplication \cite{malik_homomorphic_2023}. Simply, two encrypted numbers can be added together and the numbers are never revealed in plaintext. After adding the two encrypted numbers, the sum can be decrypted to derive the results. 

Wikipedia is a resource that demonstrates the Paillier encryption schemes \cite{wikipedia_paillier_2023}. Tracing through the calculations is obtainable without the use of HE libraries. Although, math libraries will be needed.

\subsubsection{Pyfhel HE}
FHE schemes are written in C++ but there is a scheme that is written in Python. Ibarrondo et al. introduced Pyfhel, an open source encryption library. It is a useful tool for learning FHE \cite{ibarrondo_pyfhel_2021}. The library is based on a python wrapper for the C++ Microsoft SEAL Backend and implements the BFV algorithm for integers and CKKS for operations using real numbers \cite{Catalfamo_Ahomo_2022}. Microsoft SEAL is an open source homomorphic encryption library that allows math computations on encrypted data \cite{sealcrypto}. With either Pyfhel or Microsoft SEAL, users may conduct unlimited addition and multiplication operations on encrypted data. The noise, which previously prevented unlimited operations on HE schemes, is controlled and may be observed. Afterward, decryption should reveal accurate results in plaintext. Examples of Pyfhel operations may be found at \cite{ibarrondo_ibarrondpyfhel_2023}.

\section{Conclusion}
The use of Cloud services is on the rise along with the need to protect data. Searchable Encryption (SE) is a method used to protect data with its encryption as it allows operations on such data. Plain-text is never exposed in an untrusted server. Research into SE continues. SE is extends into many branches and is a complex process. As research continues in many areas, we view its functionality.  

\begin{figure}
    \centering
    \includegraphics[width=0.90\linewidth]{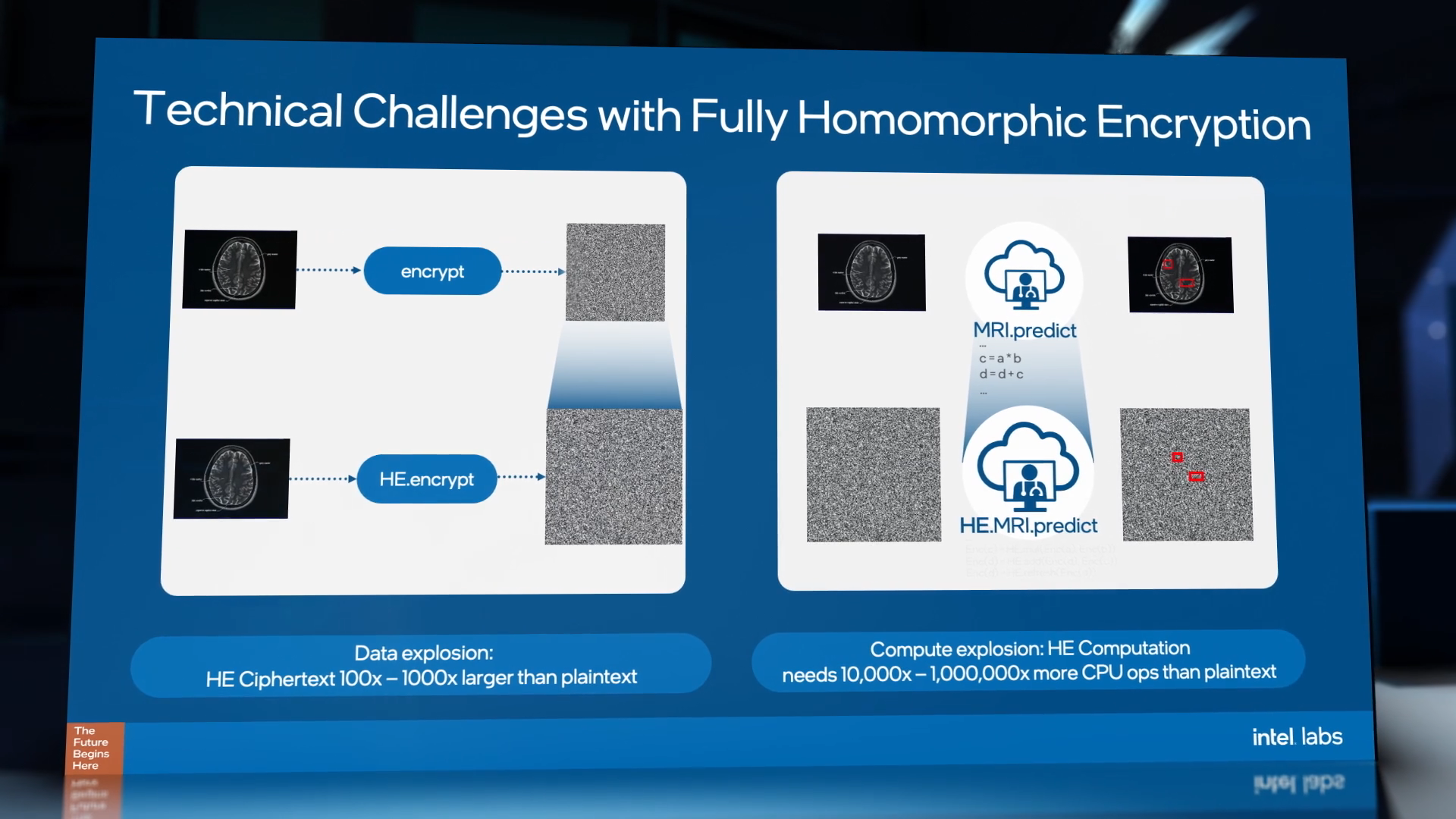}
    \caption{Internet Article claims that FHE is a million times slower than plaintext schemes.}
    \label{fig:FHEarticle}
\end{figure}

Nowadays, researchers have addressed various functionalities and combined them to make SE applicable to a wide variety of situations. Multiple words, fuzzy words, range and spatial searches, verifiable searches, and other functions and tools such as indexes, bloom filters, trees, etc are now used in conjunction with each other. Fully homomorphic encryption (FHE) is the latest type of SE. In his dissertation, Dr. Gentry made the first functional model in 2009. Since then, it has further developed and expanded. Intensive computing may be involved but research has increased its viability. Although, we may see articles claim that Fully Homomorphic Encryption is not a viable scheme. See figure \ref{fig:FHEarticle}

Public articles may not be peer-reviewed but we should objectively evaluate them. As stated at the beginning of this paper, \emph{Intel to Build Silicon for Homomorphic Encryption} is an Internet article. It states that FHE is a million times slower, compared to plaintext operations, and not useful \cite{cutress_intel_2021}. Depending on the dataset and type of operations, this could slow computer processing. Datasets, such as MRI, are highly complex especially when encrypted. Artificial Intelligence has the ability to separate patient data from MRI images. The MRI images do not need to be encrypted before they are analyzed. Only patient information needs to be removed or hidden for protection. Furthermore, research should consider the social justification of working without encrypted data. 

Overall, we are progressing when it comes to searching encrypted data. Current research improvements are not focusing on a single function such as solely looking at keywords, trees, indexes, etc. Improvement are made by using a combination of functionalities. Research should continue by incorporating most if not all pertinent functions to make it as useful as Internet searches. It is worthwhile to continue the research in the encryption field as we cannot downplay the need for data privacy.

\bibliographystyle{abbrvnat}
%\bibliographystyle{unsrtnat}
%\bibliography{main}  %%% Uncomment this line and comment out the ``thebibliography'' section below to use the external .bib file (using bibtex) .


\end{document}